\documentclass[twocolumn,showpacs,preprintnumbers,amsmath,amssymb]{revtex4}
\usepackage{graphicx,psfrag}
\usepackage{dcolumn}
\usepackage{bm}

\begin{document}

\preprint{YITP-05-??}

\title{Overdamping Phenomena near the Critical Point 
in O($N$) Model}
\author{Kazuaki Ohnishi}
 \email{kohnishi@yukawa.kyoto-u.ac.jp}
\author{Teiji Kunihiro}
 \email{kunihiro@yukawa.kyoto-u.ac.jp}
\affiliation{Yukawa Institute for Theoretical Physics, Kyoto University,
Kyoto 606-8502, Japan}
\date{\today}

\begin{abstract}
We consider the dynamic critical behavior of the propagating mode
for the order parameter fluctuation of the O($N$) Ginzburg-Landau
theory, involving the canonical momentum as a degree of freedom. We
reexamine the renormalization group analysis of the Langevin equation
for the propagating mode. We find the fixed point for the propagating
mode as well as that for the diffusive one, the former of which is
unstable to the latter. This indicates that the propagating mode
becomes overdamped near the critical point. We thus can have a
sufficient understanding of the phonon mode in the structural phase
transition of solids. We also discuss the implication for the chiral
phase transition.
\end{abstract}

\maketitle

The Quantum Chromodynamics (QCD) is believed to have a rich phase
structure in the temperature ($T$) and baryon chemical potential
($\mu_{\text{B}}$) plane \cite{Hatsuda:1994pi,Rajagopal:2000wf}.
The study of the dynamic ({\it i.e.}, nonequilibrium) critical
phenomena of the second order QCD phase transitions including the
chiral phase transition
\cite{Hatsuda:1984jm,Wilczek:1992sf,Boyanovsky:2001pa,Son:2001ff,
Koide:2003ax,Ohnishi:2004eb},
the tricritical point, the critical end point
\cite{Berdnikov:1999ph,Scavenius:2000qd,Fujii:2003bz,Son:2004iv,Koide:2004ph}
and the transitions associated with the color superconductivity
\cite{Kitazawa:2001ft} is not only of theoretical interest but also 
of fundamental importance for understanding of the relativistic heavy
ion collision experiments and the early universe where the systems
make time developments.

One of the characteristic features of the dynamic critical phenomena
is the critical slowing down, which means that it takes very long time
for the system near a critical point to relax into the equilibrium
state. The long relaxation time is attributed to the slow motion of
the long-wavelength fluctuations of the so-called slow (or gross)
variables. The slow variables are the fundamental degrees of freedom
for description of the dynamic properties near critical points. Usually
the slow variables are identified with the collection of the order
parameter and the conserved quantities of the system.

The long-wavelength fluctuations of the slow variables make up what we
call the slow modes. One should note that there are two kinds of slow
modes; the propagating mode and the diffusive mode. The propagating mode
involves oscillation with dissipation and corresponds to a pole with
both the real and imaginary parts of the spectral function for the
slow variables. The diffusive mode, on the other hand, is purely
dissipative and may be called the relaxational mode, which corresponds
to a pole with the imaginary part only.

In general, it is possible to classify critical points into the
universality classes. For the static ({\it i.e.}, equilibrium) case,
as is well known, the universality class can be determined solely by
the symmetry and dimensions of the system. For the dynamic case, a
classification scheme was proposed by Hohenberg and Halperin
\cite{Hohenberg:1977ym}: The dynamic universality classes are
dependent on what kinds of slow variable (the order parameter and
conserved quantities) are contained in the system as well as on the
symmetry and dimensions. According to the scheme, Hohenberg and
Halperin have classified the whole critical points in the condensed
matter physics in a lucid and systematic way \cite{Hohenberg:1977ym}.

The dynamic universality class of the chiral phase transition was
first discussed in Ref.\ \cite{Wilczek:1992sf}. Hohenberg and
Halperin's classification scheme tells us that the chiral phase
transition belongs to the same dynamic universality class as that of
the antiferromagnet. In Refs.\ \cite{Koide:2003ax,Ohnishi:2004eb},
however, a crucial difference between the two systems has been pointed
out. The difference is in the slow modes of the order parameter
fluctuation. In the disordered phase, the order parameter fluctuation
in the antiferromagnet is known to be a diffusive mode, while the
order parameter fluctuation in the chiral phase transition gives the
meson or particle mode, which is a propagating mode.
It has been argued \cite{Ohnishi:2004eb} that in order to describe the
meson mode appropriately, the canonical momentum conjugate to the
order parameter in addition to the order parameter itself and
conserved quantities is needed as a slow variable. The necessity
incorporating both the order parameter and its canonical momentum may
be accepted if we think of the Heisenberg equation for the
Klein-Gordon field. We note that the canonical momentum is neither the
order parameter nor a conserved quantity. This means that Hohenberg
and Halperin's classification scheme is not able to correctly classify
the dynamic universality class of the chiral phase transition and is
not a complete one at its face value.

In this Letter, we investigate the propagating mode for the order
parameter fluctuation in the O($N$) Ginzburg-Landau theory that
involves the canonical momentum as a degree of freedom. We analyze
the Langevin equation for the propagating mode by using the
renormalization group technique. For simplicity, we ignore other
possible conserved quantities. The system we will consider is the
simplest possible model.

In fact, we know a critical point in the condensed matter physics, the
soft mode for which is a propagating mode involving the canonical
momentum. It is the structural phase transition of solids
\cite{Cowley:1980}. The soft mode is a phonon mode, which is a
vibrational mode of the lattice. Experimentally, it is observed that
the phonon mode becomes overdamped, eventually turning into a
diffusive mode near the critical point. This fact may suggest that the
meson mode in the chiral transition also becomes overdamped and its
dynamic universality class reduces to that of the antiferromagnet near
the critical point as discussed in Ref.\ \cite{Wilczek:1992sf}. This
is indeed the case as we will see later. In a theoretical side, the
phonon mode was examined in Ref.\ \cite{Hohenberg:1977ym}, in which
essentially the Langevin equation for the propagating mode was
analyzed. However, the analysis will be found not to be adequate in
the sense discussed below. Our investigation will elucidate the
profound background of the overdamping phenomenon. The purpose of this
Letter is to give a reanalysis of the Langevin equation for the
propagating mode of the O($N$) Ginzburg-Landau model and to examine
the role of the canonical momentum in the critical dynamics. The model
includes the structural phase transition as an $N=1$ case and the
chiral phase transition as $N=4$. Our analysis brings us new findings
missed in the previous work, which lead to a deeper insight into the
structural transition and the chiral transition as well as into the
classification scheme of dynamic universality classes.

The Langevin equation that describes the propagating mode for the
order parameter fluctuation is given by a natural extension of that
for the Brownian particle \cite{Zwanzig:1973}:
\begin{align}
&\frac{\text{d}\pi_{i}(\vec{x},t)}{\text{d}t}
=
-\frac{\delta F[\phi]}{\delta \phi_{i}(\vec{x},t)}
-\Gamma \frac{\text{d}\phi_{i}(\vec{x},t)}{\text{d}t}
+\zeta_{i}(\vec{x},t),
\\
&\frac{\text{d}\phi_{i}(\vec{x},t)}{\text{d}t}
=
\lambda\pi_{i}(\vec{x},t),
\\
&\langle \zeta_{i}(\vec{x},t)\zeta_{j}(\vec{y},t')\rangle
=k_{\text B}T2\Gamma\delta_{ij}\delta(\vec{x}-\vec{y})\delta(t-t').
\label{FD}
\end{align}
The $\phi_{i}(\vec{x},t)$ is the $N$-component ($i=1,2,\cdots,N$)
order parameter and $\pi_{i}(\vec{x},t)$ is its canonical momentum.
The $F[\phi]$ is the usual O($N$) symmetric Ginzburg-Landau free
energy given by
\begin{equation}
F[\phi]
=\frac{1}{2}\int\text{d}^{d}x
\left[\vec{\nabla}\phi_{i}(\vec{x},t)\cdot\vec{\nabla}\phi_{i}(\vec{x},t)
+r_{0}\phi^{2}+\frac{1}{2}u\phi^{4}\right],
\end{equation}
where $d$ is the spatial dimension and $r_{0}$ and $u$ are the usual
static parameters. The dynamic parameters $\lambda$ and $\Gamma$
represent the square of the propagating velocity and the damping
constant, respectively. The $\zeta_{i}(\vec{x},t)$ is the white noise
which satisfies Eq.\ (\ref{FD}).

In the Fourier space, the Langevin equation reads
\begin{align}
&\frac{\text{d}\pi_{i\vec{k}}(t)}{\text{d}t}
=
-\left[\left(r_{0}+k^2\right)\phi_{i\vec{k}}\right.
\nonumber\\
&\left.
+\frac{u}{L^{d}}\sum_{0<k',k''<\Lambda}\phi_{j\vec{k}'}\phi_{j\vec{k}''}
\phi_{i\vec{k}-\vec{k}'-\vec{k}''}\right]
-\Gamma\frac{\text{d}\phi_{i\vec{k}}}{\text{d}t}
+\zeta_{i\vec{k}}(t),
\label{pi for propagating}\\
&\frac{\text{d}\phi_{i\vec{k}}(t)}{\text{d}t}
=
\lambda\pi_{i\vec{k}}(t),
\label{phi for propagating}\\
&\langle \zeta_{i\vec{k}}(t)\zeta_{j\vec{k}'}(t')\rangle
=k_{\text B}T2\Gamma\delta_{ij}\delta_{\vec{k},-\vec{k}'}\delta(t-t'),
\label{noise for propagating}
\end{align}
where $L^d$ is the system volume. The wavenumber of the fluctuations is
cutoff at $\Lambda$.

Our Langevin equation for the propagating mode should be in some
connection with the Langevin equation for the diffusive mode. In
general, when the friction or damping constant is very large, the
oscillation becomes overdamped and the oscillatory nature is lost. The
Langevin equation for the propagating mode should be reduced to that
for the overdamped or diffusive mode for the large damping constant.
Actually, this reduction can be proven explicitly if the nonlinear
interaction coupling $u$ is absent \cite{Ma:1976}. Consider the
Langevin equation
(\ref{pi for propagating},\ref{phi for propagating},\ref{noise for
propagating}) with $u=0$. If we take the overdamped limit by imposing
the condition $\lambda\Gamma^2\gg r_{0}+k^2$, then the canonical
momentum turns out to be the faster degree of freedom and we can
integrate it out explicitly to find the Langevin equation for the
slower degree of freedom $\phi_{i\vec{k}}(t)$;
\begin{align}
&\frac{\text{d}\phi_{i\vec{k}}(t)}{\text{d}t}
=-\gamma\frac{\delta F[\phi_{i\vec{k}}]}{\delta \phi_{i-\vec{k}}}
+\zeta'_{i\vec{k}}(t),
\\
&F[\phi_{i\vec{k}}]
=\frac{1}{2}\sum_{k<\Lambda}(r_{0}+k^2)\phi_{i\vec{k}}\phi_{i-\vec{k}},
\\
&\langle \zeta'_{i\vec{k}}(t)\zeta'_{j\vec{k}'}(t')\rangle
=k_{\text B}T2\gamma\delta_{ij}\delta_{\vec{k},-\vec{k}'}\delta(t-t'),
\end{align}
where $\zeta'_{i\vec{k}}(t)$ is the renormalized noise term and
$\gamma=1/\Gamma$ \footnote{
In the formalism of the Fokker-Planck equation, this reduction
corresponds to that from Kramer's equation to Smoluchowski equation.
See Ref.\ \cite{Hatta:2001ui}.
}.
This equation is for the diffusive mode and gives
nothing but the model A with $u=0$ in Ref.\ \cite{Hohenberg:1977ym}.

Now we apply the renormalization group to the Langevin equation for
the propagating mode, not necessarily assuming the overdamping
condition. The renormalization group analysis leading to the recursion
relation (\ref{RR_1}-\ref{RR_4}) has already been performed in Ref.\
\cite{Hohenberg:1977ym}. Although the following calculation is not
new, we will present the calculation below to make the discussion
self-contained.

The renormalization group program consists of the two procedures
\cite{Halperin:1972,Ma:1976}; (i) integrate out the short-wavelength
fluctuation  with $\Lambda/b<k<\Lambda$, and (ii) make the scale
transformation of
$\phi_{i\vec{k}}\to b^{1-\eta/2}\phi_{ib\vec{k}}(tb^{-z})$. After
the procedures, we have the recursion relation for the parameters
$(\lambda, \Gamma, r_{0}, u)$. We employ the $\epsilon$-expansion
assuming that $u$ is of order $\epsilon$, where $\epsilon=4-d$. In
this Letter, we consider only the lowest order of the expansion. The
dynamic response function for the order parameter fluctuation is given
in the form of
\begin{equation}
G(\vec{k},\omega)=\frac{1}
{-\frac{1}{\lambda}\omega^2+r_{0}+k^2-\text{i}\omega\Gamma
+\Sigma(\vec{k},\omega)},
\label{DRF}
\end{equation}
where $\Sigma(\vec{k},\omega)$ is the 1PI self-energy.

The diagrammatic rules for the perturbation are almost the same as
those for the diffusive model
(model A in Ref.\ \cite{Hohenberg:1977ym}). Only the difference is the
form of the response function. For the propagating mode, it includes
the $\omega^2$ term as seen from Eq.\ (\ref{DRF}). For the overdamped
limit where $\lambda\Gamma\gg \omega$
(or $\sqrt{\lambda}\Gamma\gg\Lambda$), the $\omega^2$ term disappears
and the response function reduces to that of the diffusive mode
\cite{Hohenberg:1977ym}. 
We note that for the particle mode, which is a propagating mode and
described by the system with $\lambda=1$, the pole position lies in
the time-like region in the $\omega$-$k$ plane. This is contrary to the
diffusive mode for which the pole is always in the space-like region.

\begin{figure}
\psfrag{p}{$\rho=\frac{\sqrt{\lambda}\Gamma}{\Lambda}$}
\psfrag{q}{$\rho=\infty$}
\psfrag{r}{$r_0$}
\psfrag{u}{$u$}
\includegraphics[width=8cm]{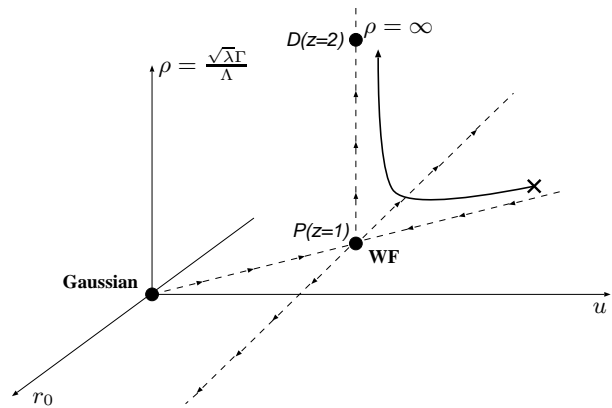}
\caption{\label{fig} The two fixed points and the
renormalization group flow.}
\end{figure}

After the integration of the fluctuation, the new parameters are
defined by
\begin{align}
r_{0}'&=\lim_{k,\omega\to 0}G^{-1}(\vec{k},\omega),\\
\Gamma'&=\lim_{k,\omega\to 0}\frac{\partial}{\partial(-\text{i}\omega)}
G^{-1}(\vec{k},\omega),\\
\frac{1}{\lambda'}&=\lim_{k,\omega\to 0}
\frac{\partial^2}{\partial(-\text{i}\omega)^2}G^{-1}(\vec{k},\omega),
\end{align}
where the last equation is absent in the diffusive model and new in
the present model.
Up to the lowest order in the $\epsilon$-expansion, the self-energy is
given by the tadpole diagram, which has no frequency and wavenumber
dependence. Thus $\Gamma'$ and $\lambda'$ receive no corrections from
the fluctuation integration. After the scale transformation, we obtain
the recursion relation;
\begin{align}
\frac{1}{\lambda'}&=b^{2-2z}\frac{1}{\lambda},
\label{RR_1}\\
\Gamma'&=b^{2-z}\Gamma,
\label{RR_2}\\
r_{0}'&=b^{2}(r_{0}+\Delta r),
\label{RR_3}\\
u'&=b^{4-d-\eta}(u+\Delta u).
\label{RR_4}
\end{align}
We note that Eq.\ (\ref{RR_1}) is given in Eq.\ (4.36) in
Ref.\ \cite{Hohenberg:1977ym}.
The first two equations are for the dynamic parameters while the last
two for the static ones.
We note that the recursions for the dynamic and static parameters are
decoupled to this order. The recursions for the static parameters
are the usual ones for the static Ginzburg-Landau theory with
$\Delta r=u(N/2+1)\int\frac{\text{d}^{d}q}{(2\pi)^{d}}(r_{0}+q^2)^{-1}$
and $\Delta u=-u^{2}(N+8)/2\int\frac{\text{d}^{d}q}{(2\pi)^{d}}
(r_{0}+q^2)^{-2}$, and give rise to the Gaussian and Wilson-Fisher (WF)
fixed points as usual. Although there are two dynamic parameters, one
of them just fixes the time scale and only some ratio of the two is
meaningful. It is useful to define the dynamic parameter as
$\rho\equiv\sqrt{\lambda}\Gamma/\Lambda$. From the recursion relation,
we find that there arise two fixed points associated with the WF fixed
point in the parameter space $(\rho, r_{0}, u)$\footnote{Actually, we
have two fixed points associated with the Gaussian fixed point as
well. The two sets of the two fixed points have the same feature with
respect to the $\rho$-direction because the recursions for the dynamic
and static parameters are decoupled. We will concentrate on the two
fixed points for the Wilson-Fisher fixed point.}. See Fig.\ \ref{fig}.

The two fixed points are located at $\rho=0$ and $\rho=\infty$, which
will be denoted as P and D, respectively. The fixed point P has the
dynamic critical exponent $z=1$ \cite{Boyanovsky:2001pa, Ohnishi:2004eb}
and represents the purely propagating mode, while the fixed point D
has the dynamic exponent $z=2$ \cite{Halperin:1972} and
corresponds to the diffusive mode. The dynamic exponents are obtained
in a usual manner \cite{Halperin:1972,Ma:1976}. The arrows indicate
the flow of the renormalization group. We see that the parameter
$\rho$ is relevant with respect to the fixed point P and the line
extended from the Gaussian fixed point to the WF fixed point gives the
critical `surface.' Thus the fixed point P is unstable to the fixed
point D. The system represented by the point near the critical
surface, such as the cross in Fig.\ \ref{fig}, is firstly taken to the
vicinity of P but eventually driven away to the fixed point D; namely,
there occurs a crossover between the two fixed points.

An illuminative example of a crossover phenomenon between two fixed
points is provided us with the static critical phenomena in the magnetic
systems \cite{Cardy:1996}. Consider the Heisenberg ferromagnet which
has the rotational O(3) symmetry. Its critical property would be
controlled by the associated fixed point, {\it i.e.}, the Heisenberg
fixed point. The full rotational symmetry may be broken into the
uniaxial one by, say, distortion of the lattice, the effect of which
is described by the appropriate anisotropic Hamiltonian with the
coupling constant $g$. It is known \cite{Cardy:1996} that the
Heisenberg fixed point is unstable with respect to the parameter
$g$. The stable fixed points are supplied by the Ising or XY fixed
points, depending on the sign of $g$. A point with small $g$ in the
parameter space is driven by the renormalization group to the Ising or
XY fixed point via the vicinity of the Heisenberg fixed point. This
means that the system firstly exhibits the critical behavior of the
Heisenberg universality class, but the Ising- or XY-like behavior
eventually shows up as the critical point is approached.

Our analysis indicates that the propagating mode crosses over into
the diffusive mode near the critical point: At first, the propagating
mode is softened with the exponent $z=1$ under the control of the
propagating fixed point, and then it becomes overdamped to be governed
by the diffusive fixed point with $z=2$.

Now we have obtained a lucid explanation for the phonon mode in
the structural phase transition \footnote{
Although we have considered only the order parameter fluctuation, and
the other conserved quantity, {\it i.e.}, the energy, has not been
taken care of in this Letter, we believe that our analysis is
sufficient for the qualitative feature of the phonon mode. This would
be so because the instability of the propagating fixed point would not
be affected even if conserved quantities are taken into account.}:
The phonon mode changes its behavior from
the propagating into the diffusive as a consequence of the crossover
between the two fixed points. As we mentioned, the recursion relation
(\ref{RR_1}-\ref{RR_4})
was already derived in Ref.\ \cite{Hohenberg:1977ym}. However it was
utilized only to discuss the stability of the diffusive fixed point
with respect to the small perturbation in the $\rho$ direction. Namely
only the vicinity of the diffusive fixed point was investigated. In
the present analysis, we have examined the whole region in the
parameter space within the validity of the $\epsilon$-expansion, and
found out the unstable propagating fixed point.

The finding of the unstable propagating fixed point brings us new
insights into the structural phase transition: We can have a simple
understanding of the overdamping phenomenon as a crossover between the
two fixed points. The mechanism of how the propagating behavior
changes into the diffusive one has become more concrete. It is clear
that because of the instability of the fixed point, the propagating
mode inevitably becomes overdamped even if we start from anywhere in
the parameter space except in the $\rho=0$ plane. Moreover the
universality of the propagating behavior of the phonon mode are
confirmed by the fixed point. These ingredients give a full
explanation in terms of the renormalization group language, which is
sufficient for understanding of the phonon mode in the structural
phase transition \cite{Bausch:1978}.

We note that it is not until the propagating fixed point is given that
its universality is guaranteed. In fact, the renormalization group
analysis of the phonon mode has been restricted only to the diffusive
behavior near the critical point up to the present. The propagating
fixed point we have found gives us a firm basis to discuss the
universal nature of the propagating behavior of the phonon mode.

It should be noted that though being unstable, the propagating fixed
point gives one dynamic universality class in the sense that fixed
points and universality classes make a one-to-one correspondence. The
Heisenberg fixed point in the magnetic system actually gives one
universality class although it is unstable against the anisotropic
interaction. As can be seen from the recursion relation
(\ref{RR_1}-\ref{RR_4}), within the $\rho =0$ plane including the
propagating fixed point, the renormalization group flow closes
itself. If the system exists which has the parameters just in the
plane, the propagating fixed point is no longer unstable
\footnote{The plane within which the renormalization group flow closes
itself coincides with the $\rho =0$ plane in the leading order of the
$\epsilon$-expansion. In the higher orders, however, the plane may be
lifted from the $\rho =0$ plane, as suggested by Eq.\ (4.80) in
Ref.\ \cite{Hohenberg:1977ym}. This means that those propagating
modes, which lie in the plane and do not become overdamped, can have a
finite width.}.
If the system takes the parameters very close to the plane initially,
the overdamped region near the critical point is correspondingly small
and the critical behavior is almost governed by the propagating fixed
point. Thus the propagating behavior constitutes one dynamic
universality class.

Since the propagating mode eventually reduces to the diffusive mode
near the critical point, the canonical momentum becomes a rapid degree
of freedom and fades out of the member of the slow variables at that
point. Thus if one restricts one's interest to the very vicinity of
the critical point, Hohenberg and Halperin's classification scheme of
dynamic universality classes, which is based on the order parameter
and conserved quantities, practically works. However, the canonical
momentum is still an important degree of freedom to find the dynamic
universality class associated with the propagating fixed point. In
this respect, Hohenberg and Halperin's classification scheme may be
regarded as incomplete.

We now turn to the chiral phase transition. Our analysis predicts that
as the system approaches the critical point, the meson mode turns into
the diffusive mode after the softening \footnote{This means that the
pole moves from the time-like to the space-like region.}.
Thus the dynamic universality class of the chiral transition certainly
reduces to that of the antiferromagnet as argued in Ref.\
\cite{Wilczek:1992sf}. However the propagating behavior of the meson
mode is still important to analyze since it shows universal properties
belonging to a fixed point. There have been various approaches, such
as the mode coupling theory \cite{Koide:2003ax,Ohnishi:2004eb}, the
Nambu--Jona-Lasinio model \cite{Hatsuda:1984jm} and the microscopic
approach in Ref.\ \cite{Boyanovsky:2001pa}. Unfortunately, the
overdamping phenomenon was not noticed in these analyses within the
adopted approximations.
Actually, up to now there have been two kinds of works based on
different standpoints: Some works treat the meson mode as a diffusive
mode while others assume the mode to be propagating. These two
standpoints are reconciled now that the crossover between the two
kinds of modes has been realized. We should note that the overdamping
of the meson mode itself has a significant physical meaning that the
sigma mesons and pions are not able to propagate and lose a particle
nature near the chiral phase transition.

To summarize, we have found the fixed point for the propagating mode,
which is unstable to the fixed point for the diffusive mode.
This means that the propagating mode for the O($N$) Ginzburg-Landau
theory becomes overdamped near the critical point.
The analysis gives a satisfactory account of the character change of
the phonon mode in the structural phase transition and also predicts
the fate of the meson mode near the chiral phase transition. In the
future work, we will investigate the higher order calculation
in the $\epsilon$-expansion, which will clarify the deviations of the
dynamic critical exponents from the mean field values obtained in this
work. The $1/N$ expansion is also interesting. Moreover it is
necessary to take account of the conserved quantities for the proper
description of the structural phase transition, the chiral phase
transition and so on. We hope to report the progresses in these
directions in the future.

\begin{acknowledgments}
We are grateful to Takao Ohta for valuable comments and to  Dam T.\ Son
and H.\ Fujii for critical and valuable comments.
T.\ K.\ is supported by Grant-in-Aide for Scientific Research by
Monbu-Kagaku-sho (No.\ 14540263).
This work is supported in part by a Grant-in-Aid for the 21st Century 
COE ``Center for Diversity and Universality in Physics.''
\end{acknowledgments}


\end{document}